\documentclass[prl,twocolumn,showpacs,superscriptaddress,preprintnumbers,amsmath,amssymb]{revtex4}

\usepackage{graphicx}
\usepackage{dcolumn}
\usepackage{bm}
\usepackage[colorlinks=true,linkcolor=blue,urlcolor=blue,citecolor=blue]{hyperref}

\begin{document}

\title{Collective nature of spin excitations in superconducting cuprates probed by resonant inelastic x-ray scattering}

\author{M. Minola}
\affiliation{Max-Planck-Institut f\"{u}r Festk\"{o}rperforschung, Heisenbergstr. 1, 70569 Stuttgart, Germany}

\author{G. Dellea}
\affiliation{CNISM, CNR-SPIN and Dipartimento di Fisica, Politecnico di Milano, 20133 Milano, Italy}

\author{H. Gretarsson}
\affiliation{Max-Planck-Institut f\"{u}r Festk\"{o}rperforschung, Heisenbergstr. 1, 70569 Stuttgart, Germany}

\author{Y. Y. Peng}
\affiliation{CNISM, CNR-SPIN and Dipartimento di Fisica, Politecnico di Milano, 20133 Milano, Italy}

\author{Y. Lu}
\affiliation{Max-Planck-Institut f\"{u}r Festk\"{o}rperforschung, Heisenbergstr. 1, 70569 Stuttgart, Germany}

\author{J. Porras}
\affiliation{Max-Planck-Institut f\"{u}r Festk\"{o}rperforschung, Heisenbergstr. 1, 70569 Stuttgart, Germany}

\author{T. Loew}
\affiliation{Max-Planck-Institut f\"{u}r Festk\"{o}rperforschung, Heisenbergstr. 1, 70569 Stuttgart, Germany}

\author{F. Yakhou}
\affiliation{European Synchrotron Radiation Facility, 71 Avenue des Martyrs, Grenoble F-38043, France}

\author{N. B. Brookes}
\affiliation{European Synchrotron Radiation Facility, 71 Avenue des Martyrs, Grenoble F-38043, France}

\author{Y. B. Huang}
\affiliation{Swiss Light Source, Paul Scherrer Institut, CH-5232 Villigen PSI, Switzerland}
\affiliation{Beijing National Laboratory for Condensed Matter Physics and Institute of Physics, Chinese Academy of Sciences, Beijing 100190, China}

\author{J. Pelliciari}
\affiliation{Swiss Light Source, Paul Scherrer Institut, CH-5232 Villigen PSI, Switzerland}

\author{T. Schmitt}
\affiliation{Swiss Light Source, Paul Scherrer Institut, CH-5232 Villigen PSI, Switzerland}

\author{G. Ghiringhelli}
\affiliation{CNISM, CNR-SPIN and Dipartimento di Fisica, Politecnico di Milano, 20133 Milano, Italy}

\author{B. Keimer}
\affiliation{Max-Planck-Institut f\"{u}r Festk\"{o}rperforschung, Heisenbergstr. 1, 70569 Stuttgart, Germany}

\author{L. Braicovich}
\affiliation{CNISM, CNR-SPIN and Dipartimento di Fisica, Politecnico di Milano, 20133 Milano, Italy}

\author{M. Le Tacon}
\email[]{m.letacon@fkf.mpg.de}
\affiliation{Max-Planck-Institut f\"{u}r Festk\"{o}rperforschung, Heisenbergstr. 1, 70569 Stuttgart, Germany}

\date{\today}

\begin{abstract}
We used resonant inelastic x-ray scattering (RIXS) with and without analysis of the scattered photon polarization, to study dispersive spin excitations in the high temperature superconductor YBa$_2$Cu$_3$O$_{6+x}$ over a wide range of doping levels ($0.1 \leq x \leq 1$).
The excitation profiles were carefully monitored as the incident photon energy was detuned from the resonant condition, and the spin excitation energy was found to be independent of detuning for all $x$. These findings demonstrate that the largest fraction of the spin-flip RIXS profiles in doped cuprates arises from magnetic collective modes, rather than from incoherent particle-hole excitations as recently suggested theoretically [Benjamin \textit{et al.}  \href{http://link.aps.org/doi/10.1103/PhysRevLett.112.247002}{Phys. Rev. Lett. {\bf 112}, 247002(2014)}]. Implications for the theoretical description of the electron system in the cuprates are discussed.
\end{abstract}


\maketitle

Electronic spin fluctuations are of central importance for current models of unconventional superconductivity in $d$- and $f$-electron compounds~\cite{Scalapino_RMP2012}. Inelastic neutron scattering (INS) provides comprehensive maps of the spin fluctuation intensity at energies and momenta that are well matched to the intrinsic collective response of correlated-electron systems, and has thus played a pivotal role in motivating and guiding theoretical work on unconventional superconductors~\cite{Fujita_JPSJ2012}. Because of the limited intensity of primary neutron beams, however, INS can only be applied to materials of which large single crystals can be grown, and it is unsuitable as a probe of spin excitations in atomically thin heterostructures of complex materials, which provide perspectives for control -- and ultimately design -- of unconventional superconductivity~\cite{Hwang_NatMat2012}.

Resonant inelastic x-ray scattering (RIXS) at transition-metal $L_{2,3}$-edges has recently emerged as a powerful momentum-resolved spectroscopic probe of collective spin excitations in crystals of sub-millimeter dimensions, and in thin films and multilayers~\cite{Ament_RMP2011, Ament_PRL2009}. Recent examples of RIXS studies of spin excitations include cuprates~\cite{Braicovich_PRL2010,Braicovich_PRB2010, letacon_NatPhys2011,Dean_NatMat2013,Dean_PRB2013,Dean_PRL2013,LeTacon_PRB2013,DallaPiazza_PRB2012,Guarise_PRL2010,Schlappa_Nature2012,Schlappa_PRL2009,Dean_PRB2014}, iron-based superconductors~\cite{Zhou_NatureCommunications2013} or iridates~\cite{Kim_PRL2012}, where the intrinsic energy scale of the spin dynamics exceeds 100 meV. Initial RIXS data on the dispersion of magnons in the antiferromagnetic ``parent compounds'' of the cuprate high-temperature superconductors are fully consistent with prior INS data~\cite{Braicovich_PRL2010,letacon_NatPhys2011}. Remarkably, further RIXS studies revealed that magnon-like collective spin excitations persist in almost undiminished form even in optimally doped and overdoped cuprates, \cite{letacon_NatPhys2011,Dean_NatMat2013,Dean_PRB2013,Dean_PRL2013,LeTacon_PRB2013} where INS data are very limited. This indicates that strong electronic correlations persist even in a regime where Fermi-liquid properties have been well documented~\cite{Platé_PRL05,Vignole_Nature2008}. Motivated by these results, soft x-ray RIXS spectrometers with greatly enhanced resolution are currently under construction at many synchrotron facilities worldwide.

To realize the potential of RIXS as a probe of unconventional superconductors and other correlated-electron systems, it is imperative to develop a quantitative description of the energy- and momentum-dependent RIXS cross section. Initial theoretical work on the undoped cuprates
suggested that the RIXS intensity can be written in terms of spin-spin correlation functions~\cite{Ament_RMP2011, Ament_PRL2009, Haverkort_PRL2010}, in close analogy to the well known expressions for INS. Numerical calculations of spin excitations in the two-dimensional Hubbard model (one of the most widely used model Hamiltonians for the cuprates) for various dopings have supported this conclusion, and the results were found to be in good agreement with the magnon-like collective modes found in the RIXS experiments~\cite{Jia_NatCom2014}. On the other hand, recent theoretical work~\cite{Benjamin_PRL2014} suggested that taking into account the effect of the core hole created in the intermediate state, the RIXS data on metallic cuprates can be understood in terms of incoherent particle-hole excitations in a simple non-interacting electron picture, thus calling into question their interpretation in terms of collective modes.

Here we report Cu-$L_3$ edge RIXS experiments on the prototypical high-temperature superconductor YBa$_2$Cu$_3$O$_{6+x}$ (YBCO$_{6+x}$) that resolve this controversy. Following up on the explicit prediction of Benjamin {\it et al.}~\cite{Benjamin_PRL2014} that the RIXS profiles should shift upon detuning of the photon energy from the $L$-edge (as expected for fluorescence in a broad particle-hole continuum), we have carefully studied RIXS in both undoped and doped YBCO$_{6+x}$ as a function of photon energy. Contrary to recent RIXS experiments which appeared to support this prediction,~\cite{Guarise_NatCom2014} we find that the energy of the spin excitations is independent of the incoming photon energy at all doping levels. The magnetic signal seen in RIXS experiments therefore arises from Raman scattering from collective spin excitations, which thus remain a central ingredient for theories of unconventional superconductivity.\\

\begin{figure}[!hb] 
\includegraphics[width=0.85\columnwidth]{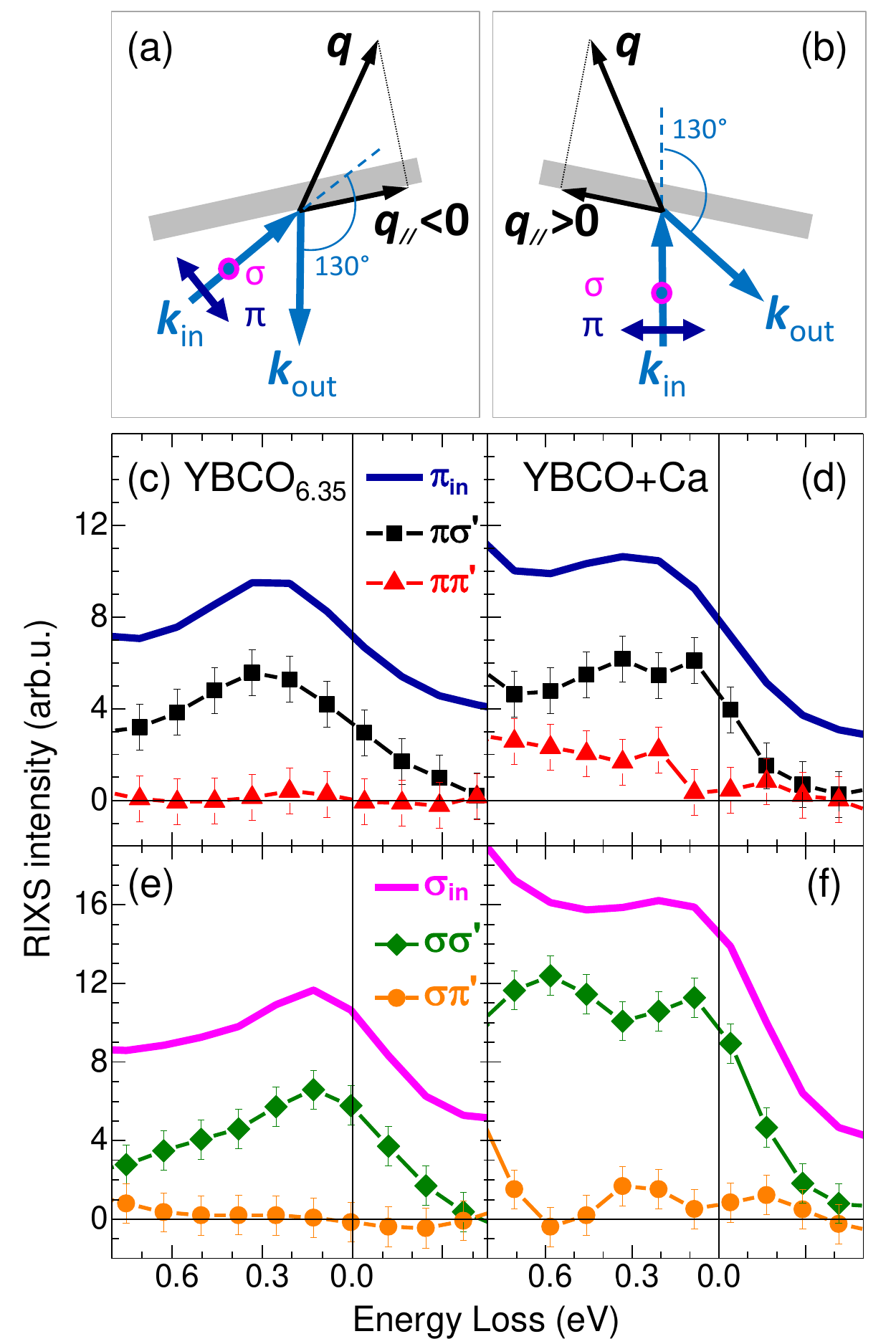}
 \caption{\label{Fig1}(Color online) Scheme of the scattering geometries allowing maximal momentum transfer with (a) grazing incident photons or (b) grazing emitted photons.
Polarization resolved measurements of underdoped YBCO$_{6.35}$ taken with incident $\pi$ (c) and $\sigma$ (e) polarized light. Polarization resolved measurements of overdoped YBCO+Ca taken with incident $\pi$ (d) and $\sigma$ (f) polarized light.}
\vspace{-5mm}
\end{figure}

The RIXS experiments were performed at the ID08 beamline of the European Synchrotron Radiation Facility (ESRF) and at the ADRESS~\cite{Strocov_JSR2010} beamline of the Swiss Light Source (SLS), using the AXES~\cite{Axes1, Axes2} and SAXES spectrometers~\cite{Ghiringhelli_RSI2006}, with respective resolutions of 350 and 160 meV.
In both cases the  scattering angle was set to 2\textit{$\theta $}=130$^{\circ}$, and the polarization of the incident photons was kept either within or perpendicular to the scattering plane ($\pi$ and $\sigma$ scattering geometry, respectively). The setup at the ESRF further allowed the analysis of the scattered light polarization~\cite{Braicovich_RSI2014}. We present systematic results obtained on various single crystals of YBCO$_{6+x}$ spanning the entire phase diagram from the nearly undoped to the overdoped regimes: $x =$ 0.1 (doping level $p \sim$ 0.014), 0.35 ($p \sim$ 0.062), 0.55 ($p \sim$ 0.114), 0.79 ($p \sim$ 0.142) to 0.99 ($p \sim$ 0.189), plus a Ca-doped sample (Y$_{0.85}$Ca$_{0.15}$Ba$_2$Cu$_3$O$_{6+x}$, $p \sim$ 0.21). Further details regarding the experimental setup as well as the crystal growth and characterization are given in the Supplemental Material~\cite{supplemental}.

\begin{figure}
\includegraphics[width=0.98\columnwidth]{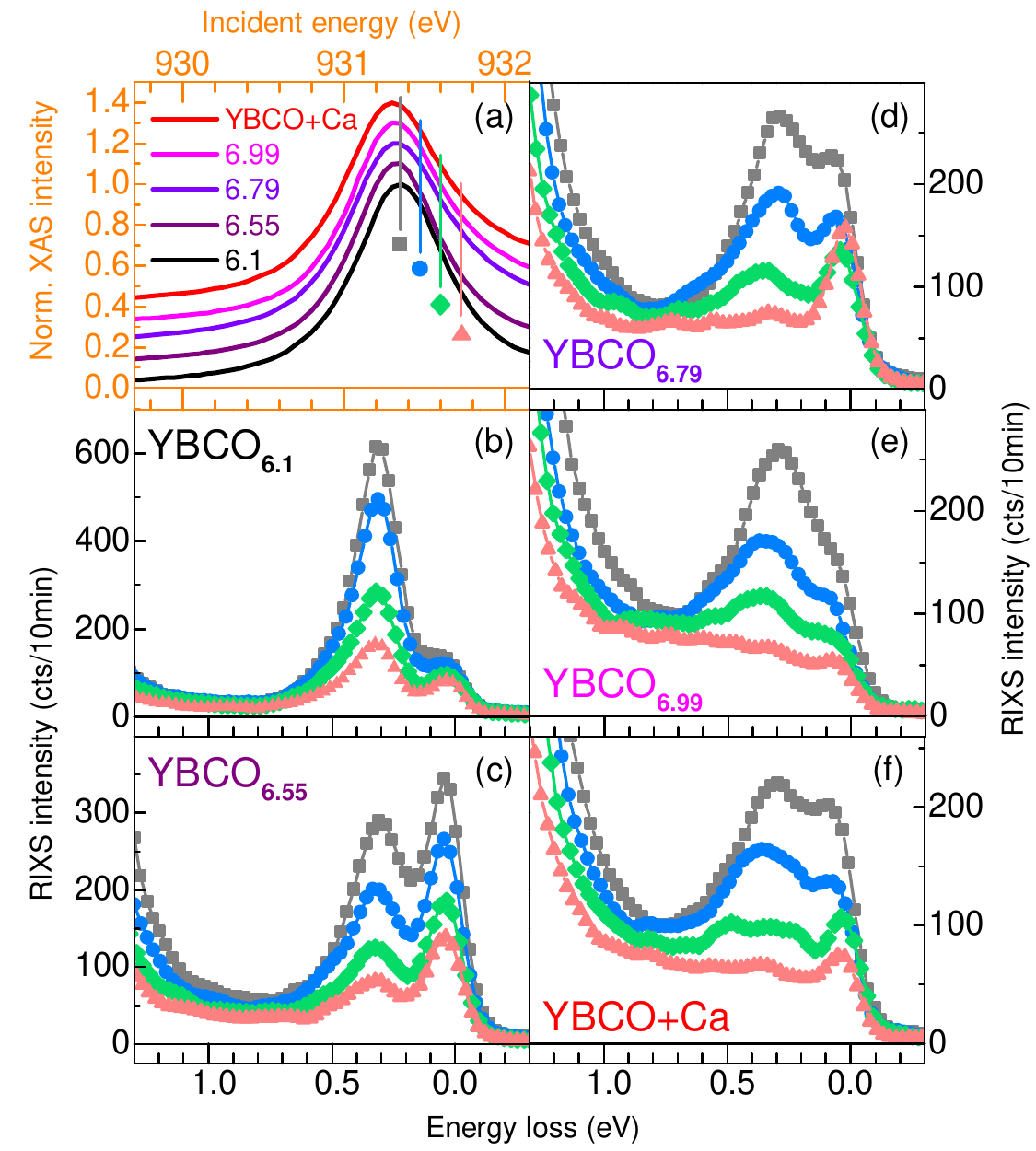}
 \caption{\label{Fig2}(Color online) Raw RIXS spectra measured with $\pi$ polarization and excited at selected incident energies across the corresponding Cu-$L_3$ edge XAS profile (a) for YBCO$_{6.1}$ (b), YBCO$_{6.55}$ (c), YBCO$_{6.79}$ (d), YBCO$_{6.99}$ (e) and YBCO+Ca (f).}
\vspace{-5mm}
\end{figure}

\begin{figure}[t]
\includegraphics[width=0.99\columnwidth]{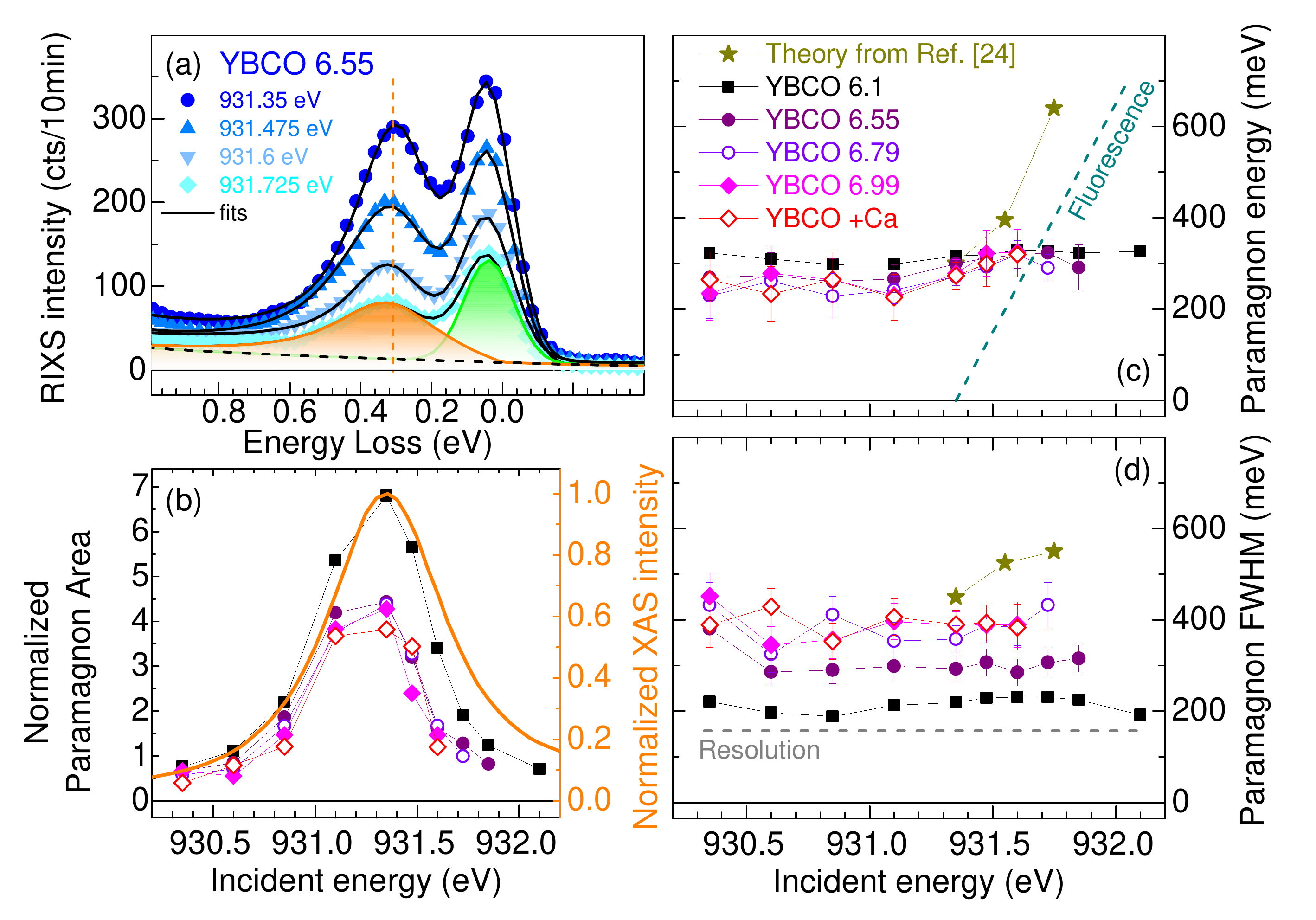}
 \caption{\label{Fig3}(Color online) (a) Examples of fitting for YBCO$_{6.55}$ excited at selected incident energies. The symbols are the raw data while the black solid lines represent the results of least-squares fits based on a Gaussian lineshape for the quasi-elastic signal (green shaded peak) and an antysimmetrized Voigt function for the paramagnon (orange shaded peak). See details in the main text. (b) Spectral weight of the paramagnon peak for all YBCO samples, as a function of incident photon energy. The magnetic integrated intensity is normalized to the area of the $dd$-excitations~\cite{Moretti_NJP2011} as excited at the L3 peak (ie the maximum of the absorption spectrum, orange solid line) and set to 100. Notice how the spectral weight is modulated by the x-ray absorption spectrum (orange solid line).
 (c) and (d) Experimental paramagnon peak position and width for all YBCO samples, as a function of incident photon energy, compared to the theoretical prediction of Ref. \onlinecite{Benjamin_PRL2014}.}
\vspace{-5mm}
\end{figure}

Before discussing the detuning experiments, we take a detailed look at the scattering geometry, which is of crucial importance for the interpretation of the resulting data. We first recall that in $L$-edge RIXS, a single spin-flip is necessarily accompanied with a change of the angular momentum of the scattered photon, resulting in a 90\r{ } rotation of the scattered beam polarization with respect to the incident one. In other words, in the experimental geometry commonly used in these experiments, where the $c$-axis lies in the scattering plane (Figs.~\ref{Fig1}-a and -b), single spin-flip events occur only in the $\pi\sigma^{\prime}$ and $\sigma\pi^{\prime}$ channels, where $\sigma^{\prime}$ and $\pi^{\prime}$ refer to the scattered x-ray polarization.
To probe high-energy magnetic excitations close to the Brillouin zone (BZ) boundary (i.e., maximal in-plane component ${\bf q}_{\parallel}$ of ${\bf q}$), one can either work with photons near grazing incidence, in which case the scattered beam is emitted close the sample surface normal ($q_{\parallel} <$ 0 in our convention~\cite{letacon_NatPhys2011,LeTacon_PRB2013,Braicovich_PRB2010,Braicovich_PRL2010} (Fig.~\ref{Fig1}-a)), or close to normal incidence with almost grazing emission ($q_{\parallel} >$ 0, Fig.~\ref{Fig1}-b). These two configurations are not equivalent. In the latter case, the $3d_{x^2-y^2} \rightarrow 2p$ transition with $\pi^{\prime}$ polarized emission is dipole-forbidden, and consequently, the RIXS signal obtained with incident $\sigma$ polarization in the scattering geometry represented in Fig.~\ref{Fig1}-b arises almost exclusively from the $\sigma\sigma^{\prime}$ channel, and does not contain single spin-flip excitations. However, with incident $\pi$ polarization, magnetic excitations are expected to account for the largest fraction of the spectral weight.

This can be experimentally demonstrated using the polarization analysis of the scattered photons, which was made possible by a recently established facility~\cite{Braicovich_RSI2014} (Figs.~\ref{Fig1}-c to f). These spectra (as all the others hereafter) were obtained with the incident photon beam at 20\r{ } from the normal incidence, resulting in an almost grazingly emitted beam (${\bf q}_{\parallel} = (0.37,0)$, close to the BZ X-point). In agreement with previous studies~\cite{letacon_NatPhys2011,LeTacon_PRB2013,Dean_NatMat2013,Dean_PRB2013,Dean_PRL2013,Braicovich_PRB2010,Braicovich_PRL2010}, the unpolarized RIXS spectra recorded with $\sigma$ polarized light on a very underdoped YBCO$_{6.35}$ reveal a strong quasi-elastic line~\cite{phonon}, whereas in the $\pi$ channel, an intense excitation is seen at around 300 meV energy loss.
Polarization analysis confirms that the signal seen in the $\sigma$ configuration arises from the $\sigma\sigma^{\prime}$ channel, \textit{i.e.} that the spin-flip fraction of the $\sigma$ spectrum is vanishingly small.
On the contrary, all the spectral weight of the spectrum recorded with $\pi$ incident photons arises from the $\pi\sigma^{\prime}$ spin-flip channel, since under these conditions the non-spin flip fraction remains negligible.
The same conclusions hold for the overdoped YBCO+Ca sample, except that a large continuum of charge excitations becomes visible in the $\sigma\sigma^{\prime}$ channel, in addition to the elastic peak. For incident $\pi$ polarization, most of the low-energy signal still originates from the $\pi\sigma^{\prime}$ channel, with a small fraction arising from the non-spin flip $\pi\pi^{\prime}$ channel.
All other scattering geometries (including the one presented in Fig.~\ref{Fig1}-a, used in Ref.~\cite{Guarise_NatCom2014}) probe a mix of spin-flip and non-spin-flip events that cannot be discriminated without systematic analysis of the scattered beam polarization.

Having firmly established $\pi$-polarized incident light and grazing emission as the optimal geometry for the investigation of magnetic excitations, we now use this geometry for high-resolution RIXS experiments without polarization analysis. Figure~\ref{Fig2} shows the dependence of the
spin-flip RIXS spectra as a function of incident photon energy for five doping levels ranging from undoped YBCO$_{6.1}$ to overdoped YBCO+Ca. Fig.~\ref{Fig2}-a shows the x-ray absorption spectra (XAS) around the Cu-$L_3$ edge for all these samples measured in total electron yield (TEY) and near normal incidence with $\pi$ polarized x-rays (the XAS intensity has been normalized to the maximum of the peak).
As previously reported~\cite{letacon_NatPhys2011,LeTacon_PRB2013}, in the doped YBCO$_{6.55}$ (Fig.~\ref{Fig2}-c), YBCO$_{6.79}$ (Fig.~\ref{Fig2}-d), YBCO$_{6.99}$ (Fig.~\ref{Fig2}-e), and YBCO+Ca (Fig.~\ref{Fig2}-f) single crystals, that do not exhibit long range magnetic order, the magnon-like feature broadens, but remains well defined \cite{note}. Clearly, the energy of the magnetic peaks is not changing with the incident photon energy. (Note that in the Supplemental Material~\cite{supplemental} we have also considered the effect of self-absorption, which turns out not to affect these conclusions).

We have fitted all RIXS spectra acquired in the $\pi$ channel close the $L_3$ resonance with resolution-limited Gaussians for the quasi-elastic peak \cite{phonon} and an antisymmetrized Voigt function for the magnetic component, on top of a background arising from the tail of $dd$-excitations which was modeled by a Voigt fit~\cite{letacon_NatPhys2011,LeTacon_PRB2013}. Fig.~\ref{Fig3}a displays typical results of such fits, and Figs.~\ref{Fig3} b-d shows the resulting paramagnon intensity, energy and width.

\begin{figure}[htbp]
\includegraphics[width=0.98\columnwidth]{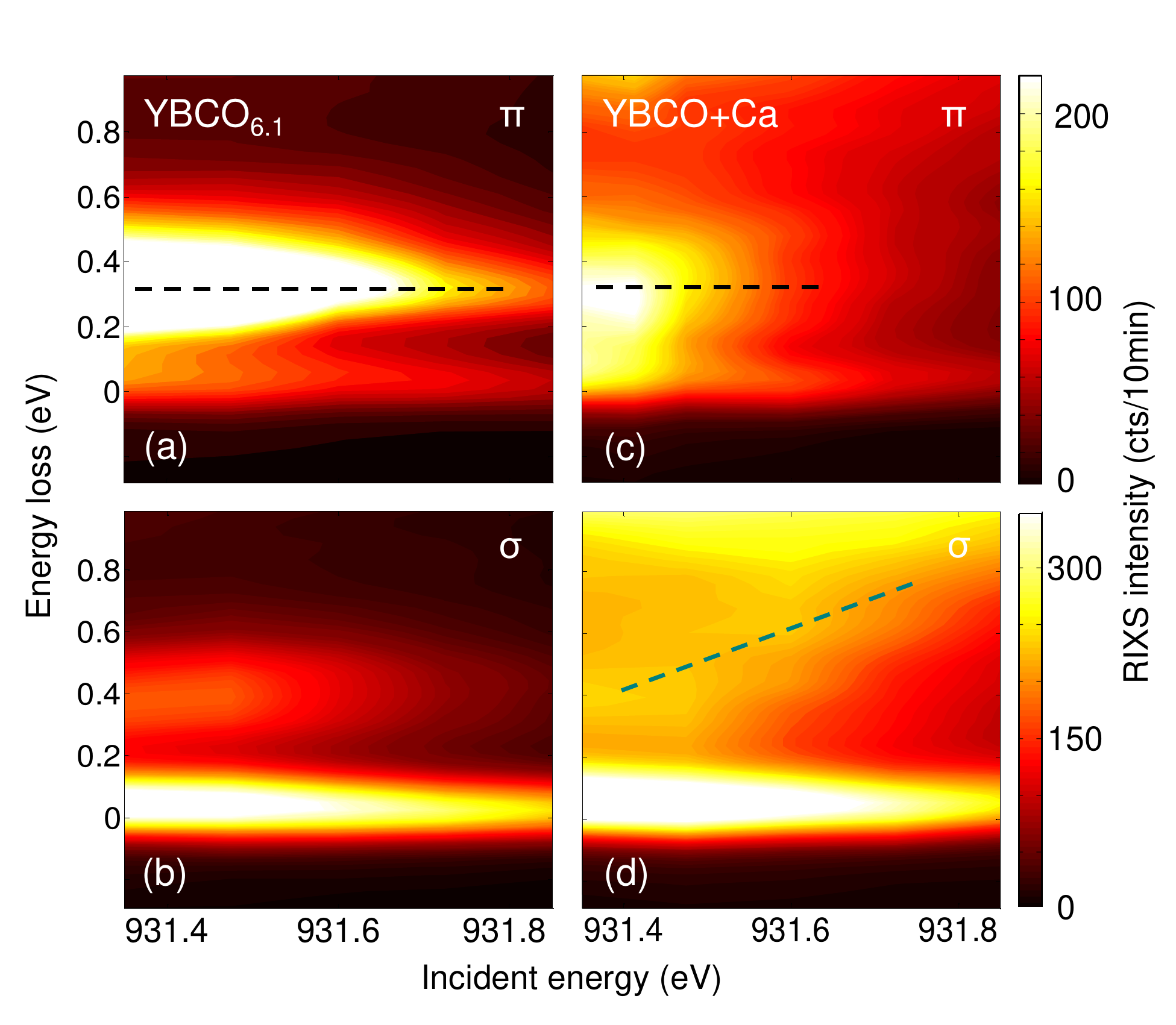}
\caption{\label{Fig4}(Color online) Color map of the energy dependence of the RIXS intensity for (a, b) antiferromagnetic YBCO$_{6.1}$ and (c, d) overdoped YBCO+Ca with $\pi$ (a, c) and $\sigma$ (b, d) polarizations. The horizontal dashed black lines highlight the energy independence of the magnetic peak position , while the inclined dashed green line is a guide to the eye underlining the fluorescence behavior of charge excitations from the doped holes.}
\vspace{-5mm}
\end{figure}

At each investigated doping level, the intensity rapidly drops as the incident photon energy is detuned from the $L_3$ edge, essentially following the x-ray absorption intensity (Fig.~\ref{Fig3}-b). The analysis further confirms that the energy and width of the magnetic peak depend at most marginally on the incident photon energy. This is in stark contrast with the behavior observed in the $\sigma$ channel shown in the color maps of Fig.~\ref{Fig4}, which display the RIXS signal as a function of the incident photon energy above the $L_3$ edge (931.35 eV) for the two extreme doping levels investigated (antiferromagnetic YBCO$_{6.1}$ and overdoped YBCO+Ca).
In the magnetically ordered compound, a stronger quasi-elastic line~\cite{phonon} accompanied with a bi-magnon signal is seen in the $\sigma$-channel. Similarly to the single spin-flip seen in the $\pi$ channel, the bi-magnon energy does not depend on the incident photon energy. In the overdoped compound with incident $\sigma$ polarization, the bimagnon cannot be resolved, but a strong continuum of charge excitations is now visible and clearly exhibits fluorescent behavior (dashed green line in Fig.~\ref{Fig4}-d).

At first glance, these results appear to be in conflict with those reported for underdoped Bi$_{1.6}$Pb$_{0.4}$Sr$_2$Ca$_{0.95}$Y$_{0.05}$Cu$_2$O$_{8+\delta}$ (at a doping level comparable with the one of our YBCO$_{6.55}$ crystal) in Ref.~\cite{Guarise_NatCom2014}. However, the incident energy dependence of the RIXS signal reported in that study originates from a mix of spin-flip and non-spin flip signal in the scattering geometry that has been used ($q_{\parallel}<$ 0, $\sigma$ incident polarization). As shown in Fig.~\ref{Fig4}-d, the non-spin flip signal originating from charge fluctuations in the $\sigma$ channel has indeed a strong fluorescent component, whereas the spin-flip one is genuinely Raman.

These observations have important consequences for the interpretation of the RIXS data. Even for doping levels as high as 20 \%, the essential component of the spin-flip response cannot be interpreted in terms of the non-interacting particle-hole excitation continuum as proposed by Benjamin \textit{et al.}~\cite{Benjamin_PRL2014}. This can be better seen in Fig.~\ref{Fig3}-c and -d, where their predictions are compared to our experimental data. One expects contributions from incoherent spin-flip excitations of the particle-hole continuum to the RIXS response on general grounds~\cite{Dean_PRB2014,Zeyher_PRB2013}. We can not rule out at this stage that such excitations indeed contribute to the background below the magnon-like collective excitations. However our data confirm that the latter constitute by far the largest fraction of the spin-flip RIXS intensity (at least up to optimal doping). A polarization analysis of the detuned RIXS spectra would allow to quantify precisely their overall contribution to the RIXS response.

Our RIXS measurements reaffirm the presence of collective spin excitations associated with short-range antiferromagnetic correlations in optimally doped and overdoped high-$T_c$ superconductors. They thus confirm the viability of the large class of models of unconventional superconductivity (including the well-known ``spin fermion'' model~\cite{Abanov_AdvPhys2003}) that are based on such correlations. Our results are also fully consistent with theories according to which the RIXS cross section can be written in terms of spin-spin correlation functions~\cite{Jia_NatCom2014,Haverkort_PRL2010}, obviating the need for materials-specific numerical simulations in the interpretation of RIXS data. With further enhancements of the energy resolution, and systematic use of the polarization analysis of the scattered photons, RIXS therefore has the potential to develop into an INS-like probe of dispersive spin excitations in a wide variety of microcrystals and thin-film structures.

We acknowledge fruitful discussion with E. Demler, I. Eremin, and A. Greco.
This work was carried out at the ADRESS beamline using the SAXES instrument jointly built by the
Paul Scherrer Institut (Villigen, Switzerland), Politecnico di Milano (Italy) and Ecole
polytechnique f\'ed\'erale de Lausanne (Switzerland).
M.M. was supported by the Alexander von Humboldt Foundation.
J. P. and T. S. acknowledge financial support through the Dysenos AG by Kabelwerke Brugg AG Holding, Fachhochschule Nordwestschweiz
and the Paul Scherrer Institut. GG, GD and YYP were supported by the PIK-POLARIX project of the Italian Ministry of Research (MIUR).


\begin{thebibliography}{}
\bibitem{phonon} Note that the maximum intensity of the quasi-elastic line is always shifted by $\sim 40$ meV from the zero-energy loss line that was determined independently by measuring a non-resonant spectrum from polycrystalline graphite. This is due to a minor phonon contribution
    that is not resolved with the present resolution.
\end{thebibliography}
\end{document}